# Controlled component segregation in vapor-deposited organic semiconductor glass mixtures

Shinian Cheng[a*], Yejung Lee[a], Lian Yu[b], Mark D. Ediger[a], Dean M. DeLongchamp[c], Camille E. Bishop[d*]

[a] Department of Chemistry, University of Wisconsin-Madison, Madison, WI 53706, United States

[b] School of Pharmacy, University of Wisconsin-Madison, Madison, WI 53705, United States

[c] National Institute of Standards and Technology, Gaithersburg, MD 20899, United States

[d] Department of Chemical Engineering and Materials Science, Wayne State University, Detroit, MI 48202, United States

*Corresponding author: chengshinian@gmail.com (S.C.); camille.bishop@wayne.edu (C.E.B)

## ABSTRACT

Multicomponent vapor-deposited organic glasses are essential in organic electronic applications, but achieving controlled component segregation at the nano/mesoscale remains a challenge, hindering the rational development of high-performance devices. In this study, we investigate binary organic semiconductor mixtures of TPD (N,N'-Bis(3-methylphenyl)-N,N'-diphenylbenzidine) and TCTA (Tris(4-carbazoyl-9-ylphenyl)amine). Despite being miscible in the bulk liquid state, the co-deposited glassy films of these two organic semiconductors exhibit a range of segregation behaviors, from homogenous to clearly phase-separated structures. We employed differential scanning calorimetry and resonant soft X-ray scattering (RSoXS) to study the component segregation behavior and used the National Institute of Standards and Technology RSoXS Simulation Suite, paired with Atomic Force Microscopy, to interpret the energy-dependent RSoXS spectra. Our results indicate that component segregation in co-deposited TPD/TCTA films is due to a kinetically-arrested nucleation-and-growth mechanism, in contrast to the segregation mechanism of a previously reported TPD/DO37 (disperse orange 37) mixture which is strongly immiscible in bulk. This work provides a demonstration of tunable molecular aggregation in organic semiconductor glasses, enabling access to a continuum of morphologies from homogeneously mixed to segregated phases.

Multicomponent organic glasses play a central role in a wide variety of established and emerging organic electronic technologies [1-5]. For example, the light-emitting layers in organic light-emitting diodes (OLEDs) are often mixtures of glassy organic semiconductors [6, 7]. While multicomponent glasses provide an effective way for fine-tuning material properties to meet specialized requirements unattainable in single-component glasses, a key challenge in the development of these multicomponent materials for specific applications is controlling their component dispersion/segregation. Understanding the mechanism of such control will be a key to optimizing glass mixtures for various technological applications, from electronics to energy devices. For example, uniform dispersion of components is preferred in OLEDs to minimize electronic interactions between emitters [8, 9], while organic photovoltaics often require mesoscale phase-segregated mixtures to facilitate charge migration pathways [10].

Physical vapor deposition (PVD) is a well-established and commercially important technique for fabricating organic glasses [11]. In this method, molecules are thermally evaporated in a high-vacuum chamber and deposited onto a temperature-controlled substrate, where the glasses are formed. PVD has been adapted to create glass mixtures by co-depositing two types of molecules onto a single substrate simultaneously. Recently, studies have shown that the properties of co-deposited glass mixtures, such as thermal stability, density, and molecular orientation, can be precisely controlled by adjusting the deposition rate, substrate temperature, and component composition [12-15]. Furthermore, recent advancements have demonstrated that PVD can also be used to control the domain size in phase-separated glass mixtures of TPD (N,N'-Bis(3-methylphenyl)-N,N'-diphenylbenzidine) and DO37 (disperse orange 37), where the two components exhibit significant immiscibility in the bulk liquid state [16]. Grazing incidence wide angle x-ray scattering measurements by Lee et al. [17] suggested that component segregation of co-deposited TCTA ((Tris(4-carbazoyl-9-ylphenyl)amine)) and $Ir(ppy)_3$ (Tris(2-phenylpyridine)iridium(III)) becomes more pronounced at higher concentrations of $Ir(ppy)_3$ and at higher substrate temperatures.

Despite recent advancements in the fundamental understanding of co-deposited organic glasses, the mechanisms governing component dispersion during co-deposition remain poorly understood. Existing studies are limited in the regime where the morphology of co-deposited glass mixtures typically mirrors their behavior in the bulk liquid state. Specifically, components that are miscible in the bulk liquid state result in well-mixed co-deposited glasses [14], while immiscible components in the bulk liquid state yield phase-separated structures in co-deposited glasses [16]. However, as non-equilibrium materials, glasses possess a nearly infinite manifold of possible packing arrangements, and their phase behavior need not be dictated by bulk equilibrium thermodynamics. A fundamental question thus arises: can we bypass equilibrium constraints to engineer phase-segregated structures from miscible components, or conversely, achieve homogenous dispersion in immiscible systems through PVD? In principle, for a given binary system, glasses that span the continuum from a homogeneous dispersion to phase-segregated structures can be prepared, independent of the phase behavior of the system at equilibrium. Achieving this level of structural control is significant not only for advancing our broader understanding of amorphous materials but also for the practical design of functional glasses. By precisely tailoring the degree of nanoscale mixing or segregation, one can strategically engineer macroscopic properties, such as optimizing charge transport pathways in organic electronics, enhancing the dissolution rates of amorphous pharmaceuticals, or fine-tuning the optical properties in photonic devices. Despite these theoretical possibilities and practical implications, experimental evidence demonstrating such a continuum of control over component dispersion in co-deposited organic glasses has not yet been presented.

In this work, we investigated glass mixtures of TPD and TCTA (Tris(4-carbazoyl-9-ylphenyl)amine), two often-used organic semiconductors in OLEDs that are miscible in bulk liquid state (Fig.1). Using differential scanning calorimetry (DSC) and resonant soft X-ray scattering (RSoXS), we examined the component segregation behavior of TPD/TCTA glasses co-deposited across a wide range of substrate temperatures and compositions. We employed the NIST RSoXS Simulation Suite (NRSS), paired with Atomic Force Microscopy (AFM), to interpret the experimental RSoXS scattering patterns. Our results demonstrate that co-deposited TPD/TCTA glasses can exhibit different degrees of component segregation, despite the two components being miscible in the bulk equilibrium liquid. Detailed analysis of the RSoXS scattering patterns indicates that the component segregation in co-deposited TPD/TCTA is governed by a kinetically-arrested nucleation and growth mechanism, producing domains with broadly distributed length scales. This behavior is fundamentally different from the previously studied co-deposited TPD/DO37 systems which is immiscible in the liquid state; in this case, spinodal decomposition controls segregation, generating domains with well-defined length scales. Our study presents definitive experimental evidence that physical vapor deposition can be used to achieve a continuum of control over component dispersion in organic glass mixtures.

## RESULTS

### 1. Bulk TPD/TCTA mixtures are miscible

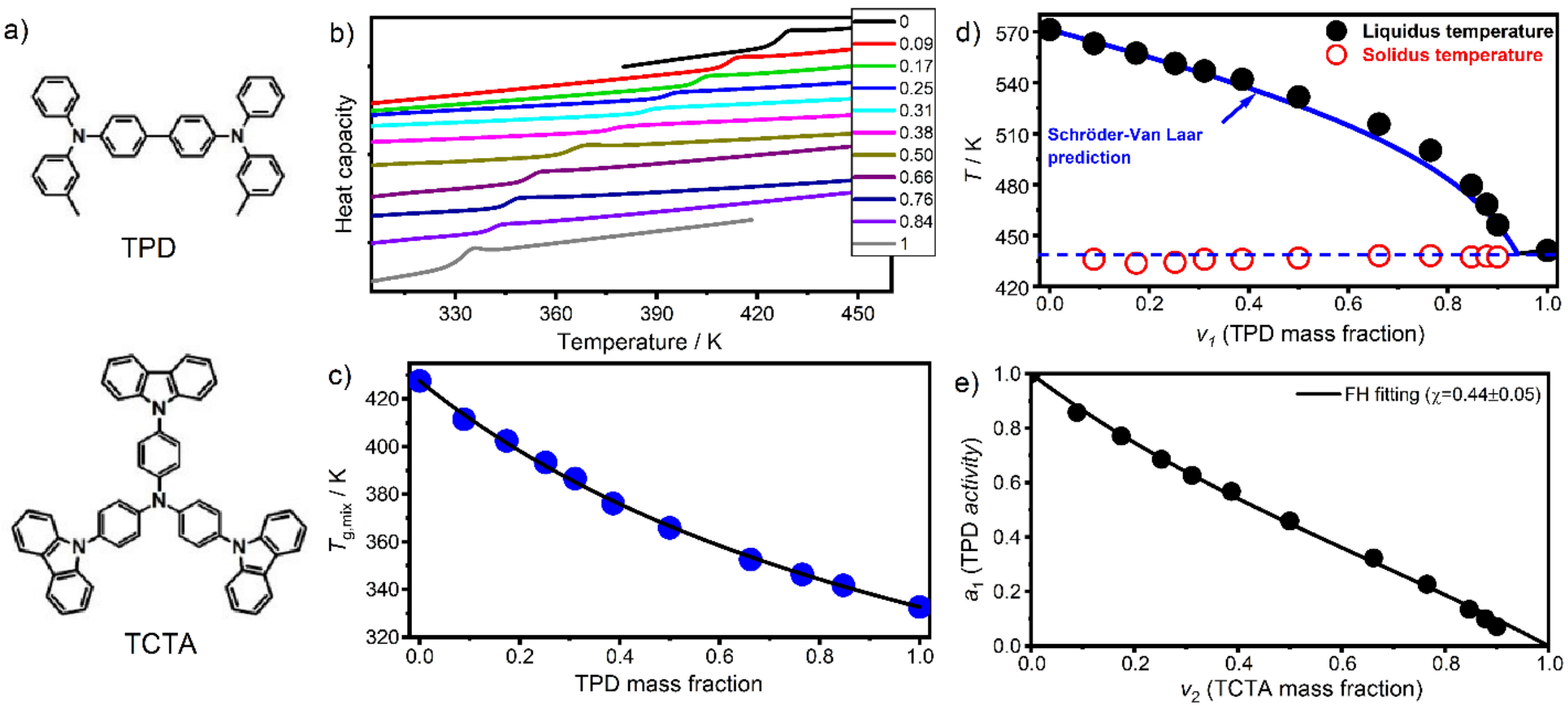

**Fig.1**: Thermograms of bulk TPD/TCTA mixtures. **a)** The chemical structure of TPD and TCTA; **b)** The DSC results of TPD/TCTA mixtures in the glass-to-liquid transition regime; the legend denotes the TPD mass fractions in the mixtures; **c)** The determined glass transition temperature of the mixtures $T_{g,mix}$ is plotted as a function of TPD mass fraction. The solid line presents the fit to the Gordon-Taylor equation; **d)** The measured liquidus and solidus temperatures from crystalline TPD/TCTA mixtures as a function of TPD mass fraction. The solid line denotes the theoretical prediction from the Schröder-Van Laar equation. The dashed line is a guide to the eyes, showing the unchanged solidus temperature with TPD mass fraction; **e)** TPD activity is plotted as a function of TCTA mass fraction and the corresponding fit to the Flory-Huggins (FH) equation with interaction parameter $\chi = 0.44 \pm 0.05$ (the solid line).

Calorimetric studies demonstrate that the bulk TPD/TCTA mixtures are miscible in their liquid states across the entire composition range. Fig.1b displays the DSC thermograms of bulk TPD/TCTA mixtures in the glass to liquid transition regime. A single glass-to-liquid transition was observed in all mixtures across the full range of TPD mass fractions from 0% (pure TCTA) to 100% (pure TPD). The determined glass transition temperature of the mixtures, $T_{g,mix}$, decreases systematically as TPD mass fraction increases and follows the Gordon-Taylor equation [18] (the black line in Fig.1c). This result suggests that homogeneous and well-dispersed glass mixtures of TPD/TCTA are formed when their bulk liquid mixture is cooled below $T_{g,mix}$. From this point, we infer that TPD and TCTA molecules are well-mixed in a single liquid phase across the entire composition range at all temperatures above $T_{g,mix}$. In this work, component compositions are quantified in terms of mass fraction, unless otherwise noted. For instance, 50%TPD/50%TCTA denotes a mixture with a composition of 50%TPD and 50%TCTA by mass.

The study of bulk crystalline TPD/TCTA mixtures (the description of sample preparation displayed in Materials and Methods) through DSC further support our inference above. For each mixture, two melting temperatures, $T_m$, are observed with the higher one being the liquidus temperature (above which the mixture is in the liquid state) and the lower one being the solidus temperature (below which both compounds in the mixture are in the crystalline state) (Fig.S1). While the liquidus temperature decreases with the increase of TPD (the low $T_m$ component) mass fraction, the solidus temperature remains almost constant across the entire composition range (Fig.1d). We compared the experimental data with the theoretical prediction of the Schröder-Van Laar equation [19] for ideal binary mixtures, where the intermolecular interactions are neutral. Clear deviations of the liquidus temperatures are observed when the TPD mass fraction ranges from 0.4 to 0.8, suggesting that the TPD/TCTA binary system is not an ideal mixture. To quantify the interaction parameter between TPD and TCTA, the TPD activity is calculated and plotted as a function of TCTA mass fraction (Fig.1e). This data is well-described by the Flory-Huggins equation [20, 21] (black line) with the interaction parameter $\chi = 0.44 \pm 0.05$. This positive interaction parameter suggests a slightly repulsive interaction between TPD and TCTA molecules. Nevertheless, based on the Flory-Huggins solution theory [22], the determined $\chi = 0.44 \pm 0.05$ is far below the estimated critical value $\chi_c = 1.77$ in TPD/TCTA mixtures (details shown in SI section 5), below which components will be miscible at equilibrium. Within the framework of the Flory-Huggins model, the much smaller $\chi$ value obtained here indicates that the thermodynamic driving force is not strong enough to result in phase separation in equilibrium liquid mixtures of TPD/TCTA. This conclusion is consistent with our inference of miscibility between TPD and TCTA above based upon a single glass transition temperature (Figs.1b and 1c).

**2. Co-deposited TPD/TCTA shows controllable component segregation**

Given the results presented above, it is surprising that component segregation is observed in co-deposited TPD/TCTA glass mixtures (Figs.2a and 2b). The temperature-dependent heat capacity of co-deposited 50%TPD/50%TCTA at $T_{sub}$ = 340 K shows two endothermic overshoot peaks, demonstrating the formation of two glassy domains with different compositions (Fig.2a, top). Since TPD has a lower $T_g$ than TCTA, the glass transition at lower temperature is assigned to TPD-rich domains. Furthermore, we observe that the component segregation behavior can be controlled by the substrate temperature. For deposition at $T_{sub}$ = 290 K, the co-deposited 50%TPD/50%TCTA exhibits a single endothermic overshoot peak, indicating the formation of a single glassy domain (Fig.2a, bottom). For this sample, the transformation onset temperature $T_{on}$ = 382 K is 1.05 times higher than the $T_{g,mix}$ = 365 K of the corresponding liquid-cooled glass mixture, highlighting the significantly enhanced stability. Given that $T_{sub}/T_{g,mix}$ = 0.8 for this sample, this high stability is consistent with the recently-established generic $T_{sub}$-dependent behavior of co-deposited organic semiconductor glasses in which no component segregation occurs [13]. It is worth noting that the DSC result for the second heating cycle of each sample, corresponding to the liquid-cooled reference glasses, shows a well-defined single glass transition with $T_{g,mix}$ = 365 K, in good agreement with that of the 50%TPD/50%TCTA bulk mixture (Figs.1b and 1c). The composition also plays a critical role in the segregation behavior of co-deposited TPD/TCTA. At $T_{sub}$ = 340 K, co-deposited glasses with 50%TPD (Fig.2a, top) and 68%TPD (Fig.2b, top), show two endothermic overshoot peaks in the glass transition region. As the TPD mass fraction decreases to 27%, a single glass transition is observed (Fig.2b, bottom). The absence of crystallization in the co-deposited TPD/TCTA mixtures is evident by the grazing incidence wide angle x-ray scattering experiments (Fig.S2), where the 2D scattering images do not show sharp rings and there are no sharp peaks observed in the integrated 1D curves.

We use the glass transition width $\Delta T$, defined as the difference between $T_{end}$ (denoting the end of the glass to liquid transition) and $T_{on}$, to qualitatively assess the degree of component segregation of co-deposited TPD/TCTA. A larger $\Delta T$ is indicative of a higher degree of component segregation. Both $T_{sub}$ and composition show clear effects on $\Delta T$ values and therefore the degree of component segregation (Fig.2c). Linear interpolation of the experimental $\Delta T$ data generates a continuous 2D color map in the rainbow color scale, changing from purple (large $\Delta T$) to red (small $\Delta T$). Mixtures that are dilute in either TPD or TCTA, corresponding to the outer contour of the map, are in the red/yellow region associated with small $\Delta T < 16$ K (*e.g.* Figs.2a and 2b, bottom). We have assigned this region as "one domain", indicating homogeneous TPD/TCTA glass mixtures. The central core of the map presents a blue/purple region where the co-deposited TPD/TCTA glasses show much large $\Delta T$ > 24 K and two glass transition regimes (*e.g.* Figs.2a and 2b, top and Figs. S3c-d). Accordingly, we assigned this region as “two domains” to indicate a high degree of component segregation. Between these two regions, a green region can be defined, where the co-deposited TPD/TCTA glasses exhibit a broad endothermic peak in DSC with 16 K < $\Delta T$ < 24 K (*e.g.* Figs.S4b-c and Figs. S4e-f). We assigned this region as “boundary”, indicating some degree of component segregation. At this point, we may conclude that the degree of component segregation for co-deposited TPD/TCTA glasses is finely tunable by substrate temperature and composition.

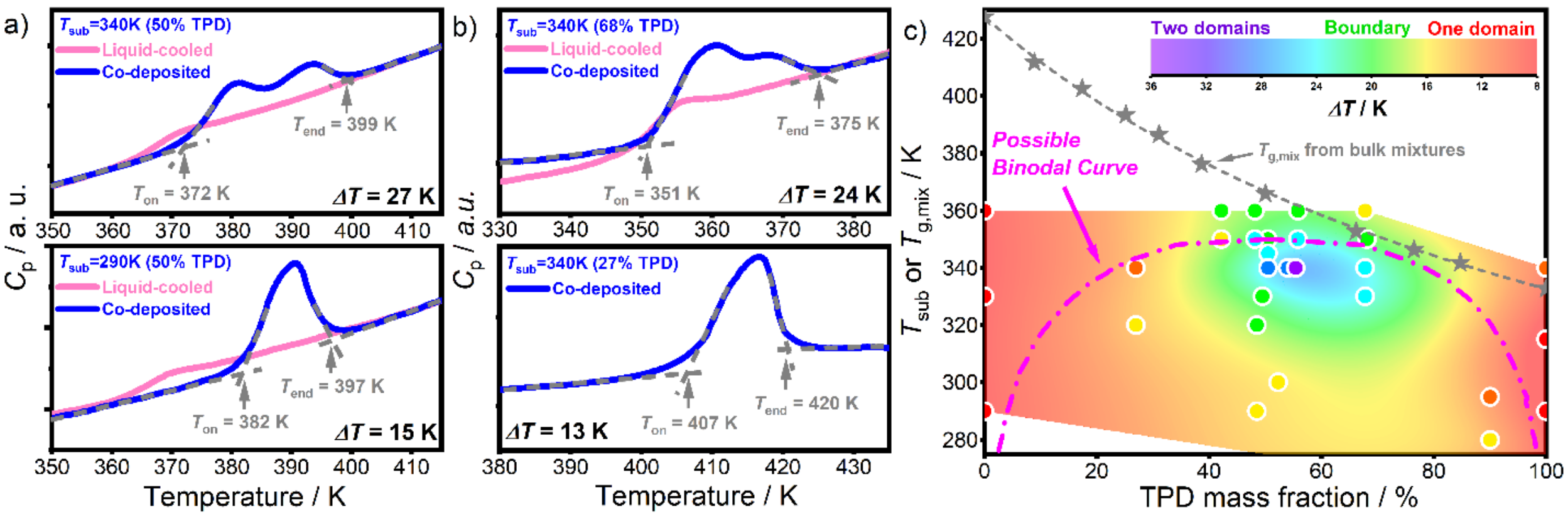

**Fig.2:** DSC characterization of component segregation for co-deposited TPD/TCTA glass mixtures. **a)** Glass transition behavior of 50%TPD/50%TCTA at $T_{sub}$ = 340 K (top) and 290 K (bottom); **b)** Glass transition behavior of co-deposited TPD/TCTA at $T_{sub}$ = 340 K with 68%TPD (top) and 27%TPD (bottom). The blue curves in a) and b) denote the co-deposited TPD/TCTA, while the pink curves denote the corresponding liquid-cooled glass. For co-deposited TPD/TCTA with 27%TPD at $T_{sub}$ = 340 K, a corresponding liquid-cooled counterpart is not available because crystallization occurred upon heating past the glass transition and the melting temperature of the formed crystal exceeds the degradation temperature of TPD. We note that this crystallization process did not occur in bulk mixtures and the underlying reasons for this different behavior are unclear. The gray dashed lines in **a)** and **b)** are guidelines to determine the onset ($T_{on}$) and end ($T_{end}$) temperatures of the glass transition region. The glass transition interval ($\Delta T$) is defined as the difference between $T_{end}$ and $T_{on}$. **c)** The variation of $\Delta T$ (characterizing the degree of component segregation) as a function of both TPD mass fraction and $T_{sub}$. The overall uncertainty of $\Delta T$ is ± 2 K, originating from the choices of guidelines to determine the $T_{on}$ and $T_{end}$. Each white-edge circle marks the temperature and composition of a given deposition, and its central color indicates $\Delta T$ of the resulting material according to the color scale in the upper right. The color map behind the white-edge circles is generated by linear interpolation, revealing the overall effects of $T_{sub}$ and composition on the degree of segregation. The gray stars present the glass transition temperatures for bulk TPD/TCTA mixtures, analyzed by the Gordon-Taylor equation (the gray dashed line), while the pink dot-dash line represents a possible binodal curve based on the FH solution theory, which is discussed below and calculated by the procedure described in SI Section 5.

**3. RSoXS measurements confirm the component segregation of co-deposited TPD/TCTA**

Resonant Soft X-ray Scattering (RSoXS) measurements, by providing robust nanoscale structural characterizations, validate the proposed component segregation above based on calorimetric experiments. We show incident-energy dependent scattering patterns for 50%TPD/50%TCTA co-deposited at $T_{sub}$ = 320 K and 340 K (Fig.3); the former is in the "boundary zone" and the latter clearly shows two domains by DSC (Fig.2c). In Figs.3a and 3b, the features of the $I$ (intensity) vs. $q$ (scattering vector magnitude) curve show important changes in shape and relative intensity at different X-ray photon energies. For clarity, we examine five energies of interest in Fig.3c. At 270.0 eV, well below the carbon K-absorption edge of 284.0 eV, the scattering from the $T_{sub}$ = 320 K sample is relatively featureless, while for $T_{sub}$ = 340 K, there is a pronounced low-$q$ ($\approx$ 0.03 nm$^{-1}$) shoulder. At the energies near and above the absorption edge, we observed that scattering for both films has a distinct fringe feature, at 0.062 nm$^{-1}$ for $T_{sub}$ = 320 K and 0.070 nm$^{-1}$ for $T_{sub}$ = 340 K. These features resemble form-factor scattering from a mixture of polydisperse spheres [23], suggesting that roughly circular or spherical forms exist in both films. Notably, the shape of the scattering does not show a pronounced Bragg-like peak as previously reported for co-deposited glasses of TPD/DO37 [16], suggesting that the phase-separated structure in the current TPD/TCTA mixture is significantly different. All RSoXS data was acquired in two X-ray polarizations of $P$=0° and 90°; negligible anisotropy was observed, allowing the energy dependence to be interpreted as entirely compositional. Once integrated over all azimuthal angles, data for the two polarizations agreed; therefore, we show data from only one polarization here. All patterns have been normalized and fluorescence-subtracted consistent with published work [24, 25]. Example two-dimensional, un-normalized, normalized, and fluorescence-subtracted data as well as anisotropy quantification are shown in SI Section 6 at all photon energies measured. Due to uncertainties in fluorescence subtraction procedures as shown in Fig. S9, we do not attempt to draw conclusions from the "high $q$" ($q$ > 0.4 nm$^{-1}$) data. We consider absorption effects negligible since our films are approximately 135 nm thick.

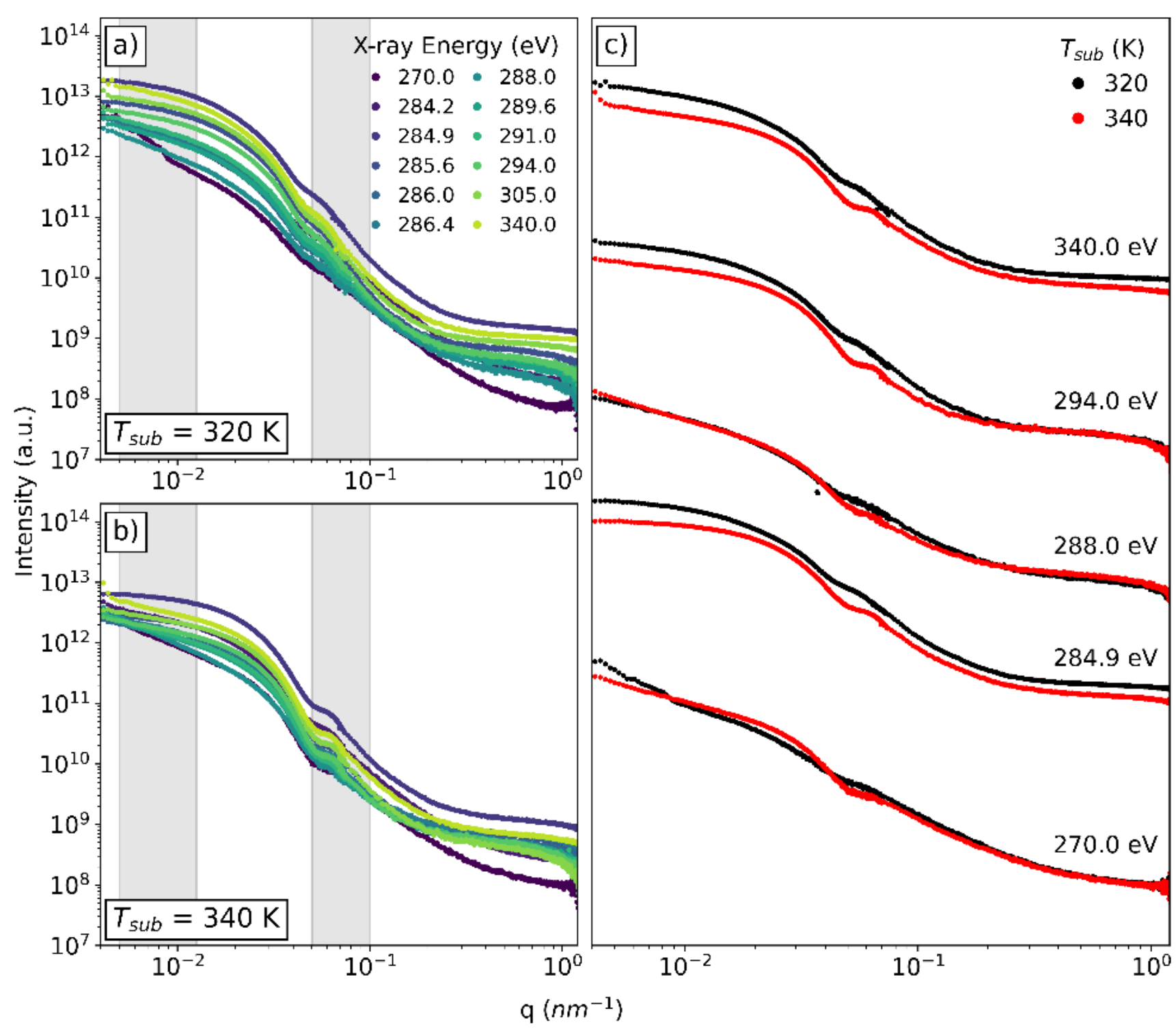

**Fig.3:** Soft X-ray scattering intensity over all scattering angles *vs* scattering vector $q$ for co-deposited 50%TPD/50%TCTA glass mixtures at $T_{\text{sub}}$ = **a)** 320 K and **b)** 340 K. Twelve energies are shown here for clarity; full data at 114 photon energies are shown in SI Section 6. Grey boxes show the main $q$-ranges of interest for integrations that are subsequently explained in Fig.4. **c)** An alternate presentation of a subset of data shown in the two previous panels, at the two $T_{\text{sub}}$ overlaid at 5 energies. Curve pairs have been vertically shifted for clarity.

The energy-dependence of the X-ray scattering intensity changes based on $T_{\text{sub}}$ of preparation. We define the integrated scattering intensity (ISI) as

$$ISI = \int_{q=q_i}^{q=q_f} I(q) \cdot q^2 \, dq \quad (1)$$

where $q_i$ and $q_f$ are the lower and upper $q$-bounds of integration. The $q$-bounds for our two areas of interest are shown as grey boxes in Figs.3a and 3b. The ISI quantifies the intensity of scattered X-rays, with the choice of $q$-bounds effectively determining the corresponding length scales of investigation. In Figs.4a and 4b, we show the integrated scattering intensity (ISI) of both films when integrated at low and medium $q$ (0.005 to 0.011 $nm^{-1}$ and 0.050 to 0.100 $nm^{-1}$, respectively). We will interpret the energy dependence of these q-bounded ISI as a rough guide to the nature of the phases contributing to contrast via correlations within a certain reciprocal space size scale, with the understanding that true deconvolution would require a full model beyond the scope of our current analysis. The low-$q$ region corresponds to length scales ca. 1 µm, while scattering signal in the higher-$q$ range will be due to objects less than 1μm in dimension. Due to uncertainties in fluorescence background subtraction, we focus primarily on the shape of the ISI dependence on energy, rather than its absolute magnitude.

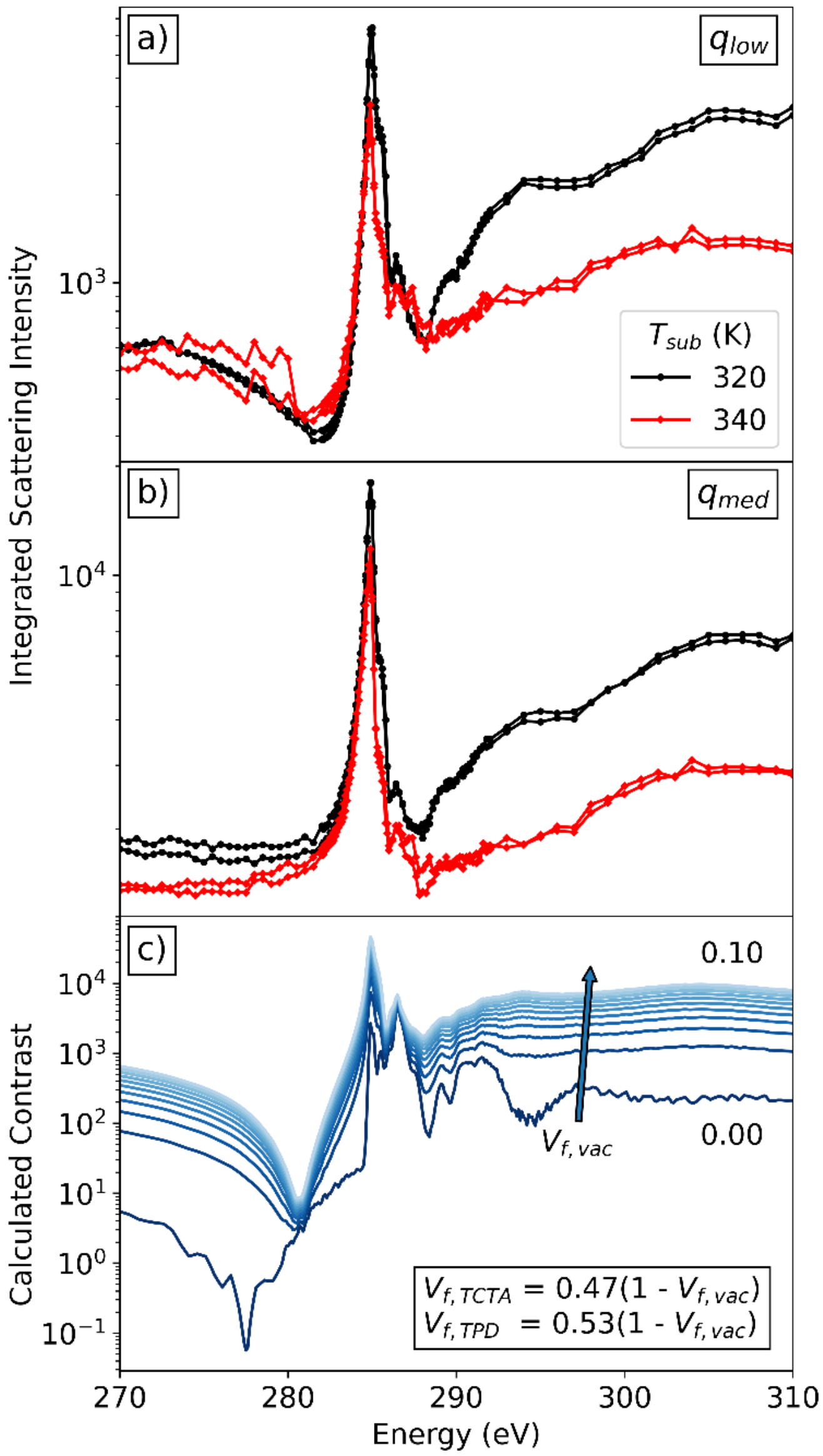

**Fig.4:** Experimental integrated scattering intensities and theoretically calculated scattering contrasts. All graphs share a common x-axis of X-ray energy. **a)** Integrated scattering intensity at "low $q$", (0.005 to 0.013 nm$^{-1}$), for the two films at all X-ray photon energies. The two traces shown for each $T_{sub}$ are from sequential measurements, one with X-ray polarization $P = 0°$, and the other with $P = 90°$. **b)** The analogous traces where the ISI integration limits are for the "medium $q$" region bounded by $q = (0.050$ to $0.100$ nm$^{-1}$). **c)** Calculated expected scattering contrasts at varied vacuum volume fraction $V_{f,vac}$ from NEXAFS spectroscopy. Calculations assume $\rho_{TPD} = 1.140$ g/cm$^3$ and $\rho_{TCTA} = 1.275$ g/cm$^3$. Formulae for $V_{f,TCTA}$ and $V_{f,TPD}$ are calculated using the densities along with the 50%/50% mass fractions.

By comparing the experimental ISI *vs*. energy to calculated scattering contrasts, we may better understand the nature of the phases from which scattering originates. Fig.4c shows the expected ternary scattering contrasts based on the ratio between three components – TPD, TCTA, and vacuum – calculated from energy-dependent optical constants. For optical constant measurement, Near-edge X-ray absorption fine spectroscopy (NEXAFS) was performed on separate, neat films of TPD and TCTA (SI Section 7). Using the Python package kkcalc [26], the NEXAFS spectra were transformed into real and imaginary components $\delta$ and $\beta$ of the X-ray refractive index $n$, given by $n = 1 - \delta + \beta i$. The spectra were first used to calculate the expected binary contrasts with the formula:

$$C_{A:B} = E^4 * ((\delta_A - \delta_B)^2 + (\beta_A - \beta_B)^2) \quad (2)$$

where $A$ and $B$ denote the two phases in the system from which contrast originates. The vacuum optical constants $\delta_{\text{vacuum}}$ and $\beta_{\text{vacuum}}$ are equal to 0. Using this relationship, we calculate the expected contrasts between TPD and TCTA, and between each of the two materials and vacuum, which are shown in SI Section 7. To calculate the ternary contrasts shown in Fig.4c, the binary contrasts are weighted by their respective volume fractions $V_f$, by:

$$C_{TPD:TCTA:vac} = V_{f,TPD}V_{f,vac}C_{TPD:vac} + V_{f,TCTA}V_{f,vac}C_{TCTA:vac} + V_{f,TPD}V_{f,TCTA}C_{TPD:TCTA} \quad (3)$$

Vacuum scattering is caused by voids in the film, either internal or due to surface roughness. Critically, the assumed density of each component has a large effect on the calculated TPD-TCTA binary contrast, since it is used to convert the NEXAFS to refractive indices. It is difficult to measure the absolute density of our amorphous films, and it is plausible that the densities in the mixtures are not the same as in neat films. As a starting point, we use published values from neat glassy films of $\rho_{\text{TCTA}} = 1.275$ g/cm$^3$ [27] and $\rho_{\text{TPD}} = 1.140$ g/cm$^3$ [28]. We calculate the expected contrasts using these two values, with possible alternate expected contrasts with density variation included in SI Section 7. Importantly, even a small contribution from vacuum contrast causes a substantial dip in ISI around 282.0 eV.

At low $q$, the pre-edge region (270.0 eV) of the ISI for both $T_{\text{sub}}$ films are strongly consistent with the inclusion of vacuum as a participating third phase, with decreasing intensity and a prominent dip going from 270.0 to 282.5 eV (Fig.3a). In contrast, the ISI vs. energy stays relatively constant in the pre-edge region of the medium-$q$ ISI region (Fig.3b). This suggests that scattering at longer length scales has a significant contribution from surface roughness or other voids in the film, while the physical origin of the medium-$q$ signal must be due to two solid phases. The peak at 284.9 eV is consistent with expected contrasts, but it provides no special insight into the participating phases since all calculated scattering contrasts are maximized at that energy. Two subtle features in the $T_{sub}$ = 340 K trace are consistent with an interpretation that it reveals slightly more contribution from TPD-TCTA binary contrast than does the $T_{sub}$ = 320 K film. The first is a diminished shoulder in scattering directly after the peak at approximately 285.4 eV, which is more consistent with the more rapid drop-off in contrast with energy in the TPD-TCTA binary contrast. The second is a peak at 287.4 eV only observed in the $T_{sub}$ = 340 K glass, which coincides with a small peak in the expected TPD-TCTA binary contrast. Anisotropy caused by molecular orientation, such as at interfaces between domains, can substantially contribute to scattering contrast [24]. To rule out the contribution of this contrast, we verified that this system does not show significant anisotropic scattering, as shown in Fig. S11. Since one dominant source of vacuum scattering in thin films is often surface roughness [29], we next investigate the surface topography as measured by AFM.

**4. AFM measurements reveal surface morphologies distinct from compositionally-separated structure**

AFM measurements reveal that co-deposited TPD/TCTA films were relatively smooth compared to their overall thickness. The height fluctuations of co-deposited 50%TPD/50%TCTA films at $T_{sub}$ = 290, 320, 330, and 340 K are no more than ± 5 nm in a 135 nm film (Fig. 5a), which are significantly less than those of co-deposited 50%TPD/50%DO37 (± 15 nm in a 70 nm film) [16]. Therefore, we expect a much smaller fraction of signal caused by vacuum scattering of the X-ray from surface roughness compared to the previously-published TPD/DO37 system [16]. We note that the exact same $T_{sub}$ = 320 and 340 K samples on silicon nitride were used for both AFM and RSoXS measurements.

Quantitative comparison of the power spectral density (PSD) transformations of AFM height data reveals that the phase-separated structure measured by RSoXS originates from component separation of TPD and TCTA, rather than the surface morphology of the co-deposited films. The PSDs were derived from the AFM height images through a Fourier transform. Each grouping of colored data in Fig.5 is composed of 2 to 5 PSDs, all of which have varied image size (a larger image will have a PSD that goes to lower $q$) and pixel size (the smaller the pixel size, the higher in $q$ range the PSD may go). The PSD curves are vertically shifted to form approximately continuous curves. Firstly, PSDs of all co-deposited 50%TPD/50%TCTA glasses show a peak at 0.023 $nm^{-1}$, corresponding to a real-space length scale of ~ 270 nm (Fig.5e). Secondly, glasses deposited at 320, 330, and 340 K show a peak at 0.19 $nm^{-1}$ (a length scale of ≈ 30 nm). While the resolution of the AFM image of the $T_{sub}$ = 290 K sample unfortunately does not show the entirety of this high-$q$ peak region, the two-dimensional image (Fig.5a) appears qualitatively consistent with those for the other three samples. These peaks do not appear in the RSoXS, suggesting that they are not due to compositional separation. Published work by Y. Jin *et. al.* [30] on vapor-deposited neat TPD shows similar AFM images, suggesting that this Marangoni-like morphology forms spontaneously during PVD even for a single component and does not depend on material separation [31].

Importantly, for these films, AFM height images are therefore not useful for measuring phase separation. Theoretically, for a perfectly homogeneous film with only surface roughness, a PSD of the AFM should have the exact same features as an RSoXS $I$ vs. $q$ profile with sufficient vacuum contrast for the same sample. Therefore, we conclude that for co-deposited TPD/TCTA, there is no connection between topography and phase separation, unlike for the previously published TPD/DO37 blend [16]. One may ask whether the phase separation is uniform through the depth of the film because there is no evidence for the structure seen in RSoXS at the free surface. There is a possibility that the composition is not uniform through the film depth, and a thin layer of a single component covers the very top. With the current data, there is no way to distinguish this; however, surface-sensitive NEXAFS, such as Total Electron Yield (TEY), could be used to measure the composition at the free surface. AFM height images are often used to measure phase separation in thin films;[32] in certain cases, it may be a reliable measure. The current analysis suggests that the AFM height images are not always a reliable indicator of composition separation, though this may be due to the vapor-deposited nature of the film as opposed to it separating by a bulk mechanism.

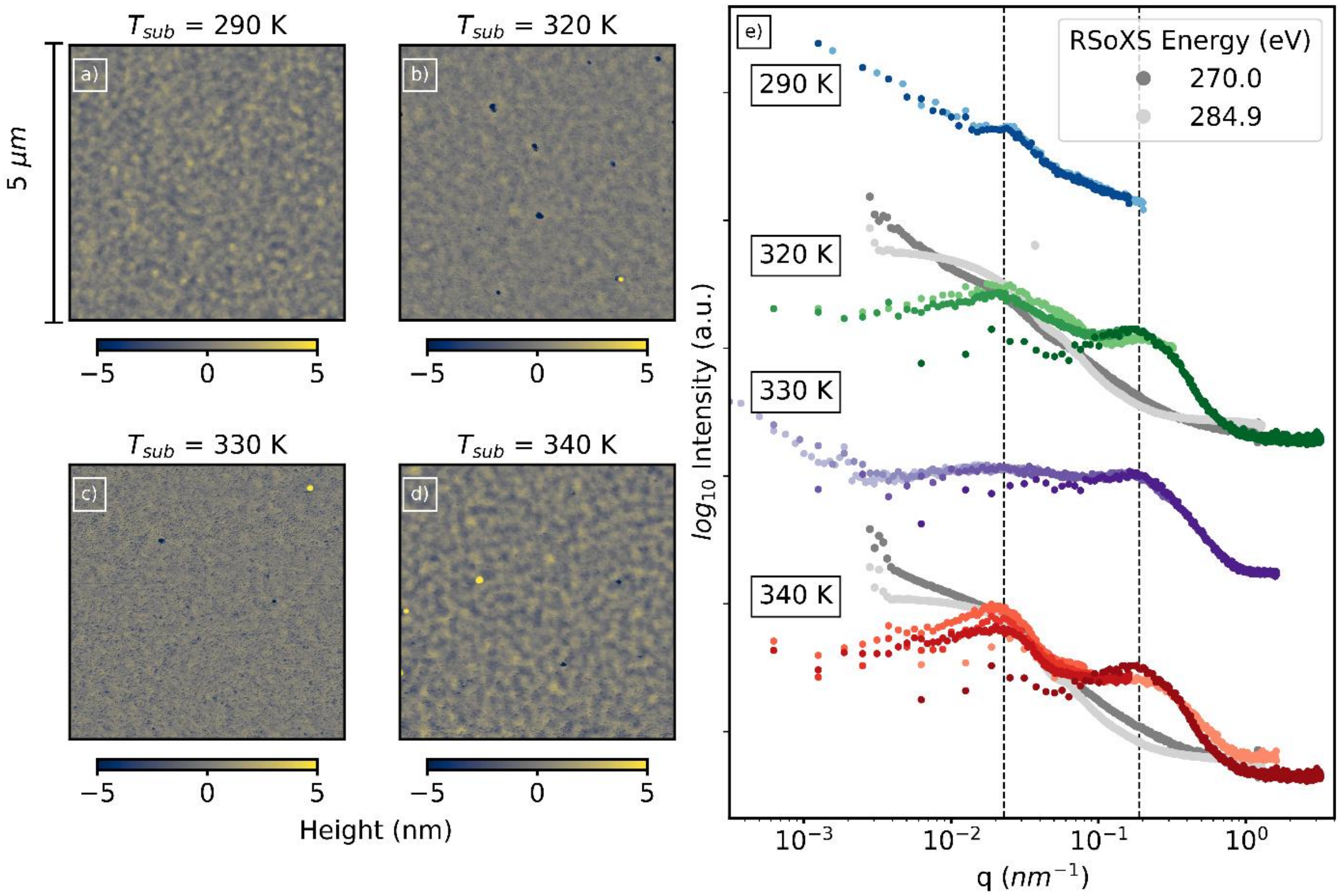

**Fig.5:** AFM reveals similar features in all samples studied. Height images for glasses deposited at $T_{sub}$= **a)** 290 K, **b)** 320 K, **c)** 330K, and **d)** 340 K. All images are 5 μm by 5 μm and are normalized to the same height intensity scale. **e)** Power spectral density (PSD) transformations of AFM data. Each distinct shade of color corresponds to a distinct AFM image; all have different q-ranges and resolutions based on image size and resolution. Each curve is vertically shifted to roughly overlay. Data shown for $T_{sub}$ = 290 K is for a thick film (~ 390 nm), while that shown for the rest of the $T_{sub}$ is on the same thin (~ 135 nm) films that were used for RSoXS. The RSoXS data presented earlier for $T_{sub}$=320 and 340K at the pre-edge (270.0 eV) and on-edge (284.9 eV) energies is underlaid for reference. Dashed lines are drawn as a guide to the eye for the two main structural features seen by the AFM.

## DISCUSSION

We begin with the Flory-Huggins solution theory[22] to understand component segregation in co-deposited TPD/TCTA. We hypothesize the existence of a "buried" binodal region below the glass transition temperature that cannot be accessed by conventional liquid cooling due to kinetic arrest. As a result, liquid-cooled TPD/TCTA mixtures exhibit a single glass transition (Fig. 1). However, PVD can access this binodal regime because of the significantly enhanced surface mobility, even at temperatures below the glass transition. Consequently, the component dispersion behavior in co-deposited TPD/TCTA is influenced by this underlying binodal. The proposed binodal curve is shown in Fig. 2c, and details of its construction are provided in SI Section 5. Qualitatively, this binodal curve accounts for the observed results. At temperatures above the binodal (*e.g*. 360 K), there is no thermodynamic driving force for segregation and co-deposition yields a homogenous glass with a narrow glass transition width. At temperatures slightly below the binodal (*e.g*. 340 K), the system thermodynamically favors two liquid phases with quite different compositions. Surface-enhanced mobility during deposition enables partial realization of these compositions. At much lower temperatures (*e.g*. 290 K), however, molecular mobility becomes too limited to allow phase separation, and co-deposition instead produces a homogenous glass despite a strong thermodynamic driving force for separation. We emphasize that the proposed binodal is intended as a qualitative framework to illustrate how temperature and composition may influence segregation in co-deposited TPD/TCTA.

We, next, turn to structural and morphological characterizations from RSoXS and AFM to further elucidate the mechanism of component segregation in co-deposited TPD/TCTA. Compared with the previously studied TPD/DO37 [16], co-deposited TPD/TCTA exhibit two notable differences: 1) the lack of well-defined, center-to-center correlation length scales in the composition separation observed by RSoXS (Fig.3) and 2) a much smoother surface observed in AFM images (Fig.5). The lack of a well-defined phase separation length scale suggests that rather than phase separating in the unstable region by spinodal decomposition, which leads to a "spinodal wavelength" and a film with significant height fluctuations [33, 34] as observed in TPD/DO37, phase separation in TPD/TCTA may occur within the metastable region of the phase diagram between the binodal and spinodal boundaries. However, it is difficult to predict these boundaries even for the bulk mixture, and there is an added complication that the surface composition does not necessarily match the bulk composition.

We hypothesize that the TPD/TCTA films grow through a two-dimensional kinetically-arrested nucleation and growth mechanism, in which the free surface layer starts evenly mixed immediately after deposition. From a purely thermodynamic perspective, Flory-Huggins theory would predict higher phase purity and more distinct separation at lower $T_{sub}$s due to the increased driving force. However, our results indicate that this trend does not always hold, and less degree of phase separation is observed at lower $T_{sub}$s, suggesting an important role of kinetics in phase separation of co-deposited TPD/TCTA. We hypothesize that the mobility is significantly reduced at the lower $T_{sub}$ so that the molecules do not have enough time at the free surface to fully separate into their equilibrium compositions before being buried by further deposition. To test our hypothesis, we use two X-ray scattering modeling approaches. The first approach is a traditional analytical expression in the dilute limit, and the second involves simulated film growth and scattering. While neither of these models exactly reproduce the experimental scattering, we reproduce the essential features of the scattering, supporting our growth hypothesis. As with any reciprocal space technique, an RSoXS scattering pattern cannot be used to directly determine the real-space structure; instead, the structure must be assumed, the theoretical scattering calculated, and then the experimental and calculated scattering are compared.

We begin with a traditional analytical model in the dilute limit, formulated to enable direct comparison with standard SAXS analysis frameworks, as illustrated schematically in Fig.6a. Our objective is to determine whether the shape of our experimental scattering $I$ vs. $q$ is consistent with the simulated scattering using the analytical model within the framework of our hypothesized 2-dimensional growth mechanisms, which would result in a distribution of disc radii. We generate log-normal distributions, shown in Fig.6b, which are consistent with a nucleation and growth mechanism [35]. The distribution of discs of TPD-rich phase A has a mean disc radius $\mu$ of 53 nm and a wider distribution than TCTA-rich phase B, which has a $\mu$ of 77 nm. As the distributions cannot be determined *a priori*, the distributions are arbitrarily generated to produce scattering consistent with the experiment; there are many distributions that could be consistent with the given data. An optimal candidate distribution is found by performing a differential evolution optimization to the experimental data. All discs are fixed to a uniform thickness of $h$=6.0nm to approximate the thickness of the region of enhanced mobility at the free surface, with orientation fixed so that the plane of the disc is oriented entirely in-the-plane of the film. The overall TPD/TCTA composition of each domain and its volume fraction are assigned based upon the $T_{on}$ of the two domains determined from the DSC data (details in SI Section 8). The composition governs the scattering contrast $\Delta\rho$, which is determined by Eq.2 using the NEXAFS-determined X-ray optical constants. We calculate the analytical expression for scattering from oriented cylinders[36] and combine the scattering from the two separate disc populations, re-weighting the scattering to be commensurate with the relative volume fraction of each phase in the sample. It should be stressed that this approach does not include any information about the organization of the domains with respect to one another; they are approximated to be in the dilute limit (in contrast to the simplified schematic of Fig.6a), with their scattering contrast calculated against a matrix of 50%TPD/50%TCTA. We provide full details of our model and calculations in SI Section 9.

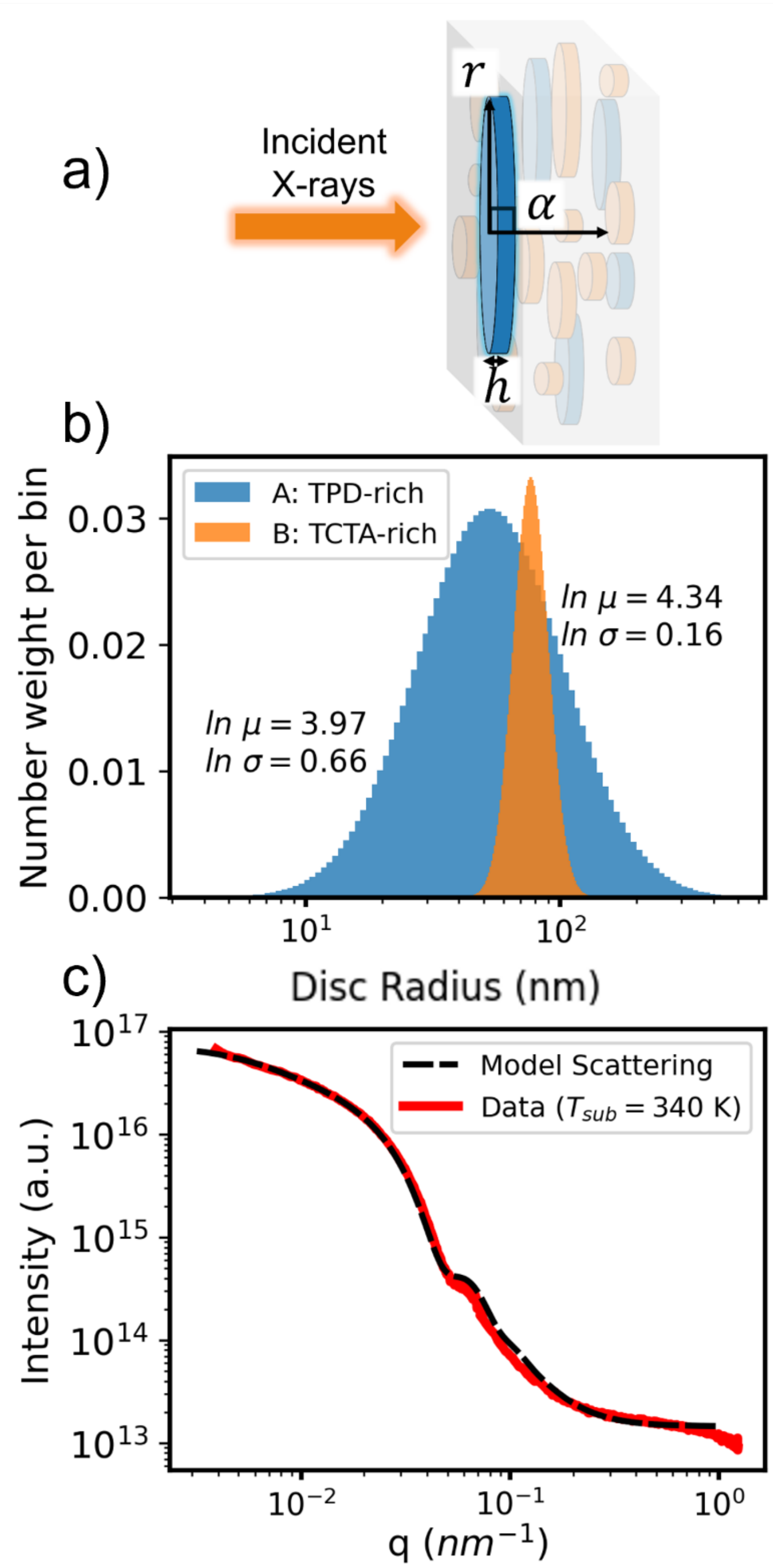

**Fig.6:** Analytical scattering expression for two populations of discs qualitatively reproduces scattering. **a)** Schematic of cylinder orientation, thickness, and radius in calculation, with the plane of the disc normal to the incident X-ray direction (and film growth direction). **b)** Disc radius distributions for the expression used to produce the simulated scattering shown in panel b. $\mu$ and $\sigma$ are the mean and standard deviation of the log-normal distribution used to generate the disc radii. **c)** A comparison of the simulated scattering from the analytical expression with the experimental RSoXS scattering data.

We find that the essential features of the experimental RSoXS data can be reproduced by analytical expressions for scattering from two populations of discs (Fig.6c). The two essential features captured are the low-q shoulder and a weak fringe at intermediate $q$. A necessary model aspect to reproduce the subtle fringe in the data, rather than distinct fringes for monodisperse spheres, is that we pick a distribution of diameters for our discs. Therefore, the analytical solution we plot, which resembles the experimental data, represents only one possible distribution of domain sizes. We stress that there are many unique distributions of disc diameters that can reproduce these essential features; since our analytical expression is already inexact, given that we are not in the dilute limit, we do not attempt to find a unique solution here.

Since our systems are not within the assumed dilute limit required to be described by the preceding analytical expression, we next use the NIST RSoXS Simulation Suite (NRSS) to simulate scattering from a more realistic system with a significant structure factor. To test our hypothesis of two-dimensional growth, we prepare a model by combining "layers" of growth into a model for simulated X-ray scattering, which we briefly describe here. A two-dimensional, single layer array of voxels corresponding to a physical size of 5000 nm (5 nm per voxel) is generated to represent the initial well-mixed phase just after deposition. As with the analytical expression, we predict the compositions of the TPD-rich and TCTA-rich domains according to the $T_{on}$s determined by DSC and determine the overall volume ratio of these two domains (SI Section 8), which appropriately set the nucleation probabilities to result in the correct overall composition. We then use a random nucleation probability model to nucleate both phases from the mixed phase, with uniform growth of the two enriched phases. We repeat this 27 times to "layer" a morphology of 27 separate 5 nm layers to prepare a 135 nm-thick film; during the nucleation and growth, each layer is grown independently from the layer below it. Finally, we use the experimental AFM results to assign vacuum fractions to give the model realistic surface roughness, which is essential for replicating experimental scattering. A scheme of the procedure is shown in Figs.7a and 7b, and the full code used to generate the model is included in the supporting information (SI Section 10).

As we increase the number of steps for each layer, we mimic enhancing the mobility at the free surface. While the initially evenly-mixed layer has enhanced surface mobility (during these timesteps), new phases with compositions determined by the DSC nucleate and grow. As shown in Fig.7b, we run the nucleation and growth for each layer for different numbers of steps. When there are more steps, it models either enhanced surface mobility caused by higher substrate temperature, or a lower deposition rate in which the surface is given more time to equilibrate. We note that deposition rate is not a factor in the present study, but could be varied to the same effect according to a rate-temperature superposition [37]. The results of the varied number of steps on the simulated scattering are shown in Fig.7c. The resonant effects are apparent based upon the different spreads of scattering intensity for the pre-edge and on-edge data.[38] We point out that this significant degree of compositional contrast occurs despite the relative impurity of "phase A" and "phase B", in which they consist only of 55% and 48% TPD in volume fraction, respectively, as estimated from the DSC $T_{on}$s data.

A key feature of our model is that we retain some mixed phase. Physically, we hypothesize that when the molecules initially deposit, they are perfectly mixed. We assume that some amount of mixed phase remains when the layer becomes immobile, since the experimental scattering retains a spherical form factor for polydisperse spheres [23]; if the enriched domains all grew into one another, this form factor would disappear, as occurs for the simulated scattering at "15 steps" shown in Fig.7c. Additionally, the presence of a mixed phase could explain the lack of two *distinct* peaks in the DSC (Fig.S4). Realistically, the mixed phase likely is not uniform and has composition gradients due to the partially hindered surface diffusion, which would result in further broadening of the glass transition range in DSC. A fully realistic model would require accounting for composition fluctuations across the surface and applying a phase diagram such as that shown in Fig.2c to every local point; however, modeling the composition gradients is beyond the scope of our model and would require more sophisticated growth modeling, perhaps through molecular dynamics or a continuum model.

Our NRSS scattering model also tests and validates hypotheses for the distinct kinetically-arrested phase separation mechanisms between the current TPD/TCTA and previously published TPD/DO37 systems [16]. As shown in Fig.7b, we test both our nucleation and growth (NG) hypothesis and a Cahn-Hilliard spinodal decomposition. The spinodal decomposition model is run in the same way as the NG model, except that instead of stochastic nucleation and growth, each layer in spinodal decomposition is given some amount of timesteps to partially phase-separate according to the deterministic Cahn-Hilliard (CH) equation [39, 40]. Our results of the two growth hypotheses (NG and CH) are shown in Fig.7d, along with the experimental RSoXS data of the currently-studied TPD/TCTA and previously-published TPD/DO37[16] in the inset. Both mechanisms are modeled with the AFM height profiles of TPD/TCTA shown in Fig.5, rather than using the previously published AFMs in Ref. [16], to test only the sub-surface separation mechanism. Both are modeled with the material optical constants for TPD and TCTA as a test for the mechanism particular to this work; therefore, we stress that the simulated CH data should not exactly match the previously published TPD/DO37 data. The CH separation mechanism (*i.e.* the spinodal decomposition) is inconsistent with the current experimental data for TPD/TCTA, as it shows a significant Bragg-like peak at $q \sim 0.1$ nm$^{-1}$ at both non-resonant and resonant energies, though it is greatly enhanced in intensity at resonance (Fig. 7d). At lower $q \sim 0.02$ nm$^{-1}$, features appear only at the resonant energy, as is seen in the previously published data of TPD/DO37 [16]. On the other hand, the NG hypothesis matches our current data of TPD/TCTA quite well. There is approximately a three order-of-magnitude enhancement in the resonant vs. non-resonant scattering intensity. Both the simulated and the experimental data show shoulders at low $q$~0.02nm$^{-1}$ that slightly change shape between the two energies. Finally, a peak at $q$~0.065nm$^{-1}$ appears only faintly in the non-resonant data, while it appears prominently at resonance. Therefore, we conclude that the growth mechanism in current phase-separated TPD-TCTA films is consistent with non-equilibrium binodal nucleation and growth. More specifically, the distribution of domain sizes and spacings from a rudimentary nucleation and growth model produces simulated scattering strongly reminiscent of the experimental data.

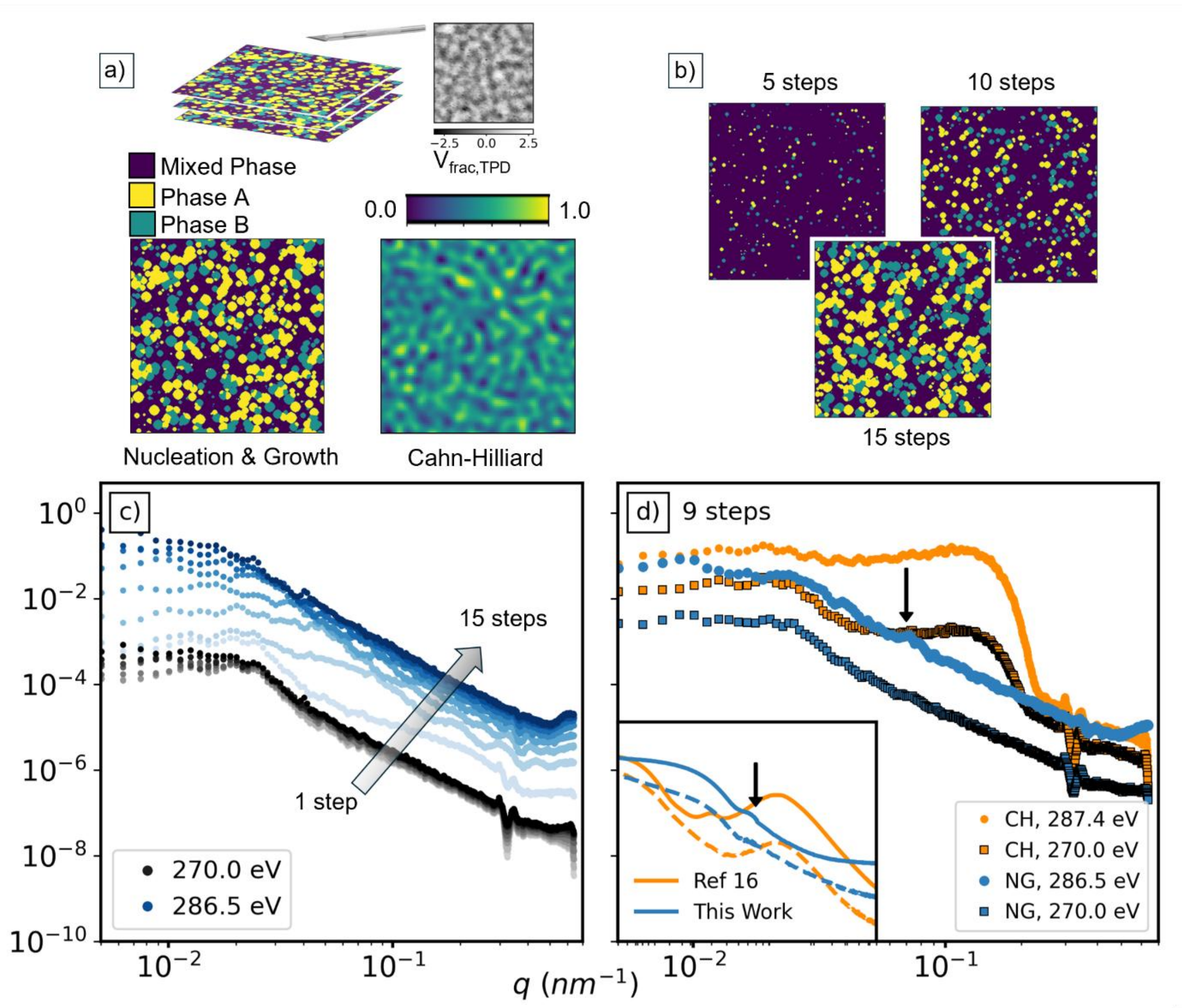

**Fig.7**. Simulated RSoXS for growth models by different mechanisms. **a)** Models are built by combining one-dimensional layers and subsequently assigning realistic surface roughness by AFM. Two hypotheses are tested – nucleation and growth of distinct phases, and a 2-dimensional Cahn-Hilliard spinodal separation. **b)** Increasing the number of steps of the simulation results in domains with greater size. **c)** Simulated scattering intensity of morphologies with each layer run for the indicated number of steps. Scattering is shown for a pre-edge (270.0 eV, greyscale) and an on-edge (287.4 eV, blue) energy. Intensity for 287.4 eV scattering is multiplied by a factor of 1.3 to fully separate the data for clarity. **d)** Simulated $I$ vs. $q$ for 9 steps of the nucleation and growth model vs. the Cahn-Hilliard decomposition. The inset shows the experimental RSoXS data of co-deposited 50%TPD/50%TCTA in current study at 270.0 eV (dashed blue) and 287.4 eV (solid blue) at $T_{sub}$ = 340 K, and previously published TPD/DO37 data at 287.4 and 285.0 eV (orange) at $T_{sub}$ = 280 K. An arrow is shown on both the main figure and the inset at 0.65 $nm^{-1}$ for comparison.

A recent report by Tsujioka *et. al* [41] suggested that molecule aggregation on a surface could happen due to Marangoni effect when two components have large difference in the surface tension, which can be characterized by the contact angle. In their work, the contact angle difference between the two components is larger than 13°. In our case, the contact angle of the two components is almost the same (Fig.S20). Considering the key to have Marangoni effect is due to the gradient of surface tension, we would exclude this effect as the main reason of the component segregation observed in co-deposited TPD/TCTA studied here.

## CONCLUSION

We investigated the component segregation behavior of binary mixtures of TPD/TCTA, two often used organic semiconductors in OLEDs. We found that despite the two components' bulk miscibility in the equilibrium liquid state, the corresponding co-deposited glassy films exhibit different degrees of component segregation, ranging from a homogeneous dispersion to clearly segregated structures, controlled by deposition temperatures and compositions. This result clearly demonstrates the applicability of physical vapor deposition to finely control the composition dispersion in organic semiconductor glass mixtures. Detailed analysis indicates that unlike previously-reported co-deposited TPD/DO37 systems forming high purity, phase-separated domains (characterized by well-defined length scale in RSoXS patterns) via spinodal decomposition, the co-deposited TPD/TCTA system studied here shows broadly distributed length scales, suggesting a compositional gradient within the formed domains. We infer that the phase separation in TPD/TCTA may occur in the metastable region, located between the binodal and spinodal boundaries and is controlled by a kinetically-arrested nucleation and growth mechanism. By analyzing the resonant scattering and performing computer simulations, we suggest that the component segregation in co-deposited TPD/TCTA system obeys a nucleation and growth mechanism. These findings significantly not only advance our understanding of amorphous materials but also provide valuable insight into designing glasses with desired properties for practical applications in photonics, pharmaceuticals, and energy storage.

## MATERIALS AND METHODS

Certain commercial equipment, instruments, or materials (or suppliers, software, etc.) are identified in this paper to foster understanding. Such identification does not imply recommendation or endorsement by the National Institute of Standards and Technology (NIST), nor does it imply that the materials or equipment identified are necessarily the best available for the purpose.

***Bulk TPD/TCTA mixture preparation*****:** As-received TPD (Sigma-Aldrich, purity ~ 99%) and TCTA (Sigma-Aldrich, purity > 97%) solid powders were combined in a clean, dry, and crack-free agate mortar with designed mass ratios. Mixing was performed manually and gently by applying consistent circular and pressing motions with the pestle for approximately 15 minutes. During mixing, material adhering to the side of the mortar was periodically scraped and reincorporated to ensure uniformity. Grinding continued until the mixture exhibited a uniform and homogenous texture.

***Physical Vapor deposited TPD/TCTA mixture preparation*****:** The co-depositions of TPD/TCTA mixtures were performed in a high-vacuum chamber with a base pressure of ~$10^{-4}$ Pa. TPD and TCTA were thermally evaporated from two independent crucibles simultaneously using resistive wire heaters and condensed on the same substrate to form glass mixtures. By controlling the deposition rates of two components individually, glass mixtures with desired mass ratios can be obtained. A quartz crystal microbalance (QCM) was applied to monitor the deposition rate and film thickness. Total deposition rate for all co-deposited glasses in this work is 0.42 ± 0.02 nm/s. The substrate temperature was held constant during deposition using a Lakeshore controller with platinum RTD sensors. To meet the requirements of various measurements, glass mixtures with different thickness were prepared on different substrates.

***Differential scanning calorimetry (DSC).*** Thermal analysis of bulk and co-deposited TPD/TCTA mixtures was performed using a TA Q2000 differential scanning calorimeter (New Castle, DE) with a scanning rate of 10 K/min for both heating and cooling processes under 50 mL/min $N_2$ purge. DSC measurements on bulk mixtures were performed using a standard aluminum pan and lid with approximately 5 mg sample sealed inside the pan for each run. DSC measurements on co-deposited films were performed using Tzero pan and lid. Co-deposited films (~ 1400 nm) onto gold foil (purchased from Barnabas Gold) were folded and loaded into the pan, and then sealed using a crimper press to achieve good contact between the test sample and the pan.

***Resonant Soft X-ray Scattering.*** RSoXS measurements of co-deposited 50%TPD/50%TCTA films (~135nm) at $T_{sub}$ = 320 to 340 K on Silicon Nitride Membranes (Norcada NX5200C) were performed at beamline SST-1 of the NSLS-II at Brookhaven National Laboratory. All samples were measured in transmission geometry with both SAXS and WAXS detectors, using approximate sample-to-detector distances of 500 and 38 mm, respectively. All data shown in this publication is taken at the carbon edge, ranging from 270 to 340 eV, with an energy resolution of 0.2 eV. Data was reduced using the PyHyperScattering Python package. SAXS and WAXS data from the same run is combined for subsequent analysis.

***Near-Edge X-ray Absorption Fine Structure Spectroscopy.*** Near-edge X-ray absorption fluorescence spectroscopy (NEXAFS) data was taken at SST-1 at the NEXAFS measurement station. NEXAFS data were collected in partial electron yield mode via channeltron detector across the Carbon K-edge with a grid bias of -150 V. Spectra were fit in QANT [42], exported, and converted to X-ray optical constants using kkcalc [26]. Incident flux was measured by a gold mesh upstream from the sample. The thickness of the films for NEXAFS measurements is around 72 nm, deposited on one-side polished silicon wafers (purchased from Virginia Semiconductor).

***Atomic Force Microscopy.*** AFM images of the same films used in RSoXS measurements were collected on a Bruker Dimension Icon in tapping mode. Scans used for modeling were taken at a resolution of 512 lines with 512 pixels per line, corresponding to a lateral resolution of 9.77 nm. Images were processed with pySPM [43]. AFM images of thicker films (~ 390 nm) co-deposited on one-side polished silicon wafers were collected using a Bruker Veeco MultiMode IV in tapping mode. The images were taken at a resolution of 256 lines with 256 pixels per line.

**Contact angle measurements**. Static contact angle measurements were performed using a DataPhysics OCA15EC measuring system to evaluate the surface wettability of vapor-deposited neat TPD ($T_{sub}$ = 296 K) and TCTA ($T_{sub}$ = 310 K) glassy films. The sessile drop method was employed at room temperature. A controlled droplet volume (~ 3 μL) of ethylene glycol was delivered using a syringe onto the film surface. Images of the droplets were captured using the integrated high-speed CCIR camera. The resulting contact angles were calculated by analyzing the droplet profile using Software dpiMAX 20P, applying a Laplace-Young fitting model to the drop baseline. For each film, three runs were applied on different regions of the film surface.

**Grazing Incidence Wide Angle X-ray Scattering (GIWAXS).** GIWAXS measurements for co-deposited 50%TPD/50%TCTA films at $T_{sub}$ = 290 K and 340 K were performed at beamline 11-3 with X-ray energy 12.7 keV at the Standford Synchrotron Radiation Lightsource. The 2D GIWAXS images were acquired at an incident angle 0.14°, higher than the critical angle for studied materials. The exposure time is 240 s for $T_{sub}$ = 290 K film and 180 s for $T_{sub}$ = 340 K film. The film thickness is 420 nm, deposited on silicon wafer. Measurements were performed in a helium atmosphere at room temperature.

**Author Contributions**

The manuscript was written through contributions of all authors. All authors have given approval to the final version of the manuscript.

**Notes**

The authors declare no competing financial interest.

## ACKNOWLEDGMENTS

S.C., L.Y., and M.D.E. acknowledge financial support from U.S. National Science Foundation through the University of Wisconsin Materials Research Science and Engineering Center (DMR-2309000). Y.L. thanks support by the U.S. Department of Energy, Office of Basic Energy Sciences, Division of Materials Sciences and Engineering (DE-SC0002161). D.M.D. was funded by the National Institute of Standards and Technology. C.E.B. acknowledges support from Wayne State University faculty startup funds and the National Research Council Research Associateship Program. This research used beamline SST-1 of the National Synchrotron Lightsource II, a U.S. Department of Energy (DOE) Office of Science User Facility operated for the DOE Office of Science by Brookhaven National Laboratory under Contract no. DE-SC001204. The use of the Stanford Synchrotron Radiation Lightsource, SLAC National Accelerator Laboratory, was supported by the U.S. Department of Energy, Office of Science, Office of Basic Energy Sciences, under Contract No. DE-AC02-76SF00515.

## REFERENCES

(1) de Zerio, A. D.; Müller, C. Glass forming acceptor alloys for highly efficient and thermally stable ternary organic solar cells. *Adv. Energy Mater.* **2018**, *8* (28), 1702741.
(2) Adhikari, T.; Rahami, Z. G.; Nunzi, J.-M.; Lebel, O. Synthesis, characterization and photovoltaic performance of novel glass-forming perylenediimide derivatives. *Org. Electron.* **2016**, *34*, 146-156.
(3) Chan, L.-H.; Lee, R.-H.; Hsieh, C.-F.; Yeh, H.-C.; Chen, C.-T. Optimization of high-performance blue organic light-emitting diodes containing tetraphenylsilane molecular glass materials. *J. Am. Chem. Soc.* **2002**, *124* (22), 6469-6479.
(4) Yokoyama, D.; Wang, Z. Q.; Pu, Y.-J.; Kobayashi, K.; Kido, J.; Hong, Z. High-efficiency simple planar heterojunction organic thin-film photovoltaics with horizontally oriented amorphous donors. *Sol. Energy Mater. Sol. Cells* **2012**, *98*, 472-475.
(5) Wu, Y.-C. M.; Molaire, M. F.; Weiss, D. S.; Angel, F. A.; DeBlase, C. R.; Fors, B. P. Synthesis of amorphous monomeric glass mixtures for organic electronic applications. *J. Org. Chem.* **2015**, *80* (24), 12740-12745.
(6) Walzer, K.; Maennig, B.; Pfeiffer, M.; Leo, K. Highly efficient organic devices based on electrically doped transport layers. *Chem. Rev.* **2007**, *107* (4), 1233-1271.
(7) Kalyani, N. T.; Dhoble, S. Organic light emitting diodes: Energy saving lighting technology—A review. *Renew. Sustain. Energy Rev.* **2012**, *16* (5), 2696-2723.
(8) Tanaka, M.; Noda, H.; Nakanotani, H.; Adachi, C. Molecular orientation of disk-shaped small molecules exhibiting thermally activated delayed fluorescence in host–guest films. *Appl. Phys. Lett.* **2020**, *116* (2).
(9) Reineke, S.; Rosenow, T. C.; Lüssem, B.; Leo, K. Improved High-Brightness Efficiency of Phosphorescent Organic LEDs Comprising Emitter Molecules with Small Permanent Dipole Moments. *Adv. Mater.* **2010**, *22* (29), 3189-3193.
(10) Moench, T.; Koerner, C.; Murawski, C.; Murawski, J.; Nikolis, V. C.; Vandewal, K.; Leo, K. Small Molecule Solar Cells. In *Molecular Devices for Solar Energy Conversion and Storage*, Tian, H., Boschloo, G., Hagfeldt, A. Eds.; Springer Singapore, 2018; pp 1-43.
(11) Mattox, D. M. *Handbook of physical vapor deposition (PVD) processing*; William Andrew, 2010.
(12) Lee, Y.; Cheng, S.; Ediger, M. High Density Two-Component Glasses of Organic Semiconductors Prepared by Physical Vapor Deposition. *J. Phys. Chem. Lett.* **2024**, *15* (31), 8085-8092.
(13) Cheng, S.; Lee, Y.; Yu, J.; Yu, L.; Ediger, M. D. Generic Behavior of Ultrastability and Anisotropic Molecular Packing in Codeposited Organic Semiconductor Glass Mixtures. *Chem. Mater.* **2024**.
(14) Cheng, S.; Lee, Y.; Yu, J.; Yu, L.; Ediger, M. D. Surface Equilibration Mechanism Controls the Stability of a Model Codeposited Glass Mixture of Organic Semiconductors. *J. Phys. Chem. Lett.* **2023**, *14* (18), 4297-4303.
(15) Lee, Y.; Cheng, S.; Yu, L.; Ediger, M. D. High kinetic stability and high density of Ir (ppy) 3 doped organic semiconductor glasses. *J. Chem. Phys.* **2025**, *163* (12).
(16) Bishop, C.; Ferron, T. J.; Fiori, M. E.; Flagg, L. Q.; Alahe, A. M. M.; Gann, E.; Jaye, C.; Bischak, C. G.; Ediger, M. D.; DeLongchamp, D. M. Resonant Soft X-ray Scattering Reveals

Hierarchical Structure in a Multicomponent Vapor-Deposited Glass. *Chem. Mater.* **2024**, *36* (18), 8588-8601.
(17) Lee, Y.; Ju, J.; Cheng, S.; Yu, J.; Yu, L.; Ediger, M. Probing component segregation and anisotropy for co-deposited glasses of TCTA and Ir (ppy) 3 by GIWAXS. *J. Chem. Phys.* **2026**, *164* (13).
(18) Gordon, M.; Taylor, J. S. Ideal copolymers and the second-order transitions of synthetic rubbers. I. Non-crystalline copolymers. *J. Appl. Chem.* **1952**, *2* (9), 493-500.
(19) Khoo, I.-C.; Wu, S.-T. *Optics and nonlinear optics of liquid crystals*; world scientific, 1993.
(20) Huggins, M. L. Solutions of long chain compounds. *J. Chem. Phys.* **1941**, *9* (5), 440-440.
(21) Flory, P. J. Thermodynamics of high polymer solutions. *J. Chem. Phys.* **1942**, *10* (1), 51-61.
(22) Lodge, P. C. H. T. P. *Polymer Chemistry*; 2007.
(23) Rieker, T.; Hanprasopwattana, A.; Datye, A.; Hubbard, P. Particle size distribution inferred from small-angle X-ray scattering and transmission electron microscopy. *Langmuir* **1999**, *15* (2), 638-641.
(24) DeLongchamp, D. M. Resonant Soft X-ray Scattering for Organic Photovoltaics. *J. Phys. Chem. B.* **2025**, *129* (13), 3529-3545.
(25) Ferron, T. J.; Fiori, M. E.; Ediger, M. D.; DeLongchamp, D. M.; Sunday, D. F. Composition dictates molecular orientation at the heterointerfaces of vapor-deposited glasses. *Jacs Au* **2023**, *3* (7), 1931-1938.
(26) Watts, B. Calculation of the Kramers-Kronig transform of X-ray spectra by a piecewise Laurent polynomial method. *Opt. Express* **2014**, *22* (19), 23628-23639.
(27) Ferron, T. J.; Thelen, J. L.; Bagchi, K.; Deng, C. T.; Gann, E.; de Pablo, J. J.; Ediger, M. D.; Sunday, D. F.; DeLongchamp, D. M. Characterization of the Interfacial Orientation and Molecular Conformation in a Glass-Forming Organic Semiconductor. *Acs Appl. Mater. Inter.* **2022**, *14* (2), 3455-3466. DOI: 10.1021/acsami.1c19948.
(28) Shibata, M.; Sakai, Y.; Yokoyama, D. Advantages and disadvantages of vacuum-deposited and spin-coated amorphous organic semiconductor films for organic light-emitting diodes. *J. Mater. Chem. C* **2015**, *3* (42), 11178-11191.
(29) Culp, T. E.; Ye, D.; Paul, M.; Roy, A.; Behr, M. J.; Jons, S.; Rosenberg, S.; Wang, C.; Gomez, E. W.; Kumar, M. Probing the internal microstructure of polyamide thin-film composite membranes using resonant soft X-ray scattering. *ACS Macro Letters* **2018**, *7* (8), 927-932.
(30) Jin, Y.; Zhang, A.; Wolf, S. E.; Govind, S.; Moore, A. R.; Zhernenkov, M.; Freychet, G.; Arabi Shamsabadi, A.; Fakhraai, Z. Glasses denser than the supercooled liquid. *Proc. Natl. Acad. Sci. U. S. A.* **2021**, *118* (31), e2100738118.
(31) Nitzsche, M. Focused Laser Spike Dewetting as a Rheology Method for Soft Matter Thin Films. **2021**.
(32) Walheim, S.; Böltau, M.; Mlynek, J.; Krausch, G.; Steiner, U. Structure formation via polymer demixing in spin-cast films. *Macromolecules* **1997**, *30* (17), 4995-5003.
(33) Böltau, M.; Walheim, S.; Mlynek, J.; Krausch, G.; Steiner, U. Surface-induced structure formation of polymer blends on patterned substrates. *Nature* **1998**, *391* (6670), 877-879.
(34) Karim, A.; Slawecki, T.; Kumar, S.; Douglas, J.; Satija, S.; Han, C.; Russell, T.; Liu, Y.; Overney, R.; Sokolov, J. Phase-separation-induced surface patterns in thin polymer blend films. *Macromolecules* **1998**, *31* (3), 857-862.
(35) Bergmann, R. B.; Bill, A. On the origin of logarithmic-normal distributions: An analytical derivation, and its application to nucleation and growth processes. *J. Cryst. Growth* **2008**, *310* (13), 3135-3138.

(36) GUINIER, A.; FOURNET, G. *SMALL-ANGLE SCATTERING OF X-RAYS*; 1955.
(37) Bishop, C.; Gujral, A.; Toney, M. F.; Yu, L.; Ediger, M. D. Vapor-deposited glass structure determined by deposition rate–substrate temperature superposition principle. *J. Phys. Chem. Lett.* **2019**, *10* (13), 3536-3542.
(38) Dalal, S. S.; Walters, D. M.; Lyubimov, I.; de Pablo, J. J.; Ediger, M. D. Tunable molecular orientation and elevated thermal stability of vapor-deposited organic semiconductors. *Proc. Natl. Acad. Sci. U.S.A.* **2015**, *112* (14), 4227-4232. DOI: 10.1073/pnas.1421042112.
(39) Izumitani, T.; Hashimoto, T. Slow spinodal decomposition in binary liquid mixtures of polymers. *J. Chem. Phys.* **1985**, *83* (7), 3694-3701.
(40) Cahn, J. W. Phase separation by spinodal decomposition in isotropic systems. *J. Chem. Phys.* **1965**, *42* (1), 93-99.
(41) Tsujioka, T.; Yamabayashi, K.; Kotani, K. Surface Glass Transition Temperature Region of Diarylethene Films Determined by Nano-Marangoni Effect. *Small* **2024**, *20* (9), 2306145.
(42) Gann, E.; McNeill, C. R.; Tadich, A.; Cowie, B. C.; Thomsen, L. Quick AS NEXAFS Tool (QANT): a program for NEXAFS loading and analysis developed at the Australian Synchrotron. *J. Synchrotron Radiat.* **2016**, *23* (1), 374-380.
(43) Scholder, O. scholi/pySPM v0.2.20 (v0.2.20). *Zenodo* **2019**. DOI: https://doi.org/10.5281/zenodo.2650457.